\documentclass[english,11pt]{article}
\usepackage[T1]{fontenc}
\usepackage[latin9]{inputenc}
\usepackage{float}

\usepackage{multicol}
\usepackage[dvipdfm]{graphicx}
\usepackage{amsmath}
\usepackage{amsfonts}
\usepackage{amssymb}
\usepackage{amsthm}
\usepackage{color}
\usepackage{babel}
\usepackage{verbatim}

\newtheorem{thm}{Theorem}
\theoremstyle{definition}

\newtheorem{lem_s}[thm]{Lemma}
\newtheorem*{cvmps_in_gamma}{CVMPS in \mbox{\boldmath$\gamma$}}

\theoremstyle{definition}

\begin{document}
\sloppy
\global\long\def\bg{BG_{n}^{\prime}}
\global\long\def\ep{EP}
\global\long\def\uep{\bigcup_{a_{i}\in p}\ep(a_{i})}
\global\long\def\uer{\bigcup_{a_{i}\in p}ER(a_{i})}
\global\long\def\use{\bigcup_{a_{i},a_{j}\in p}SE(a_{i}a_{j})}
\global\long\def\vs{\mathit{VMPSet}}

\title{Refutation of Aslam's Proof that $NP=P$}

\author{
Frank Ferraro \and Garrett Hall \and Andrew Wood \and \\
\{fferraro, ghall3, awood8\}@u.rochester.edu \and\\
Department of Computer Science\\University of
Rochester\\Rochester, NY 14627\\\\
}

\maketitle
\begin{abstract}
Aslam presents an algorithm he claims will count the number of perfect
matchings in any incomplete bipartite graph with an algorithm in the
function-computing version of $NC$, which is itself a subset of $\mathit{FP}$.
Counting perfect matchings is known to be $\#P$-complete; therefore
if Aslam's algorithm is correct, then $NP=P$. However, we show that
Aslam's algorithm does not correctly count the number of perfect matchings
and offer an incomplete bipartite graph as a concrete counter-example.
\end{abstract}

\section{Introduction}
We provide preliminary definitions from Aslam's paper, which are necessary to understanding Aslam's algorithm and what it purportedly proves. 
After presenting these 
definitions, claims and examples, we look at an overview of the
algorithm and why it purports $\mathit{NP}=P$. In Section \ref{refutations}, we refute 
some major arguments of his paper which, under his current
construction, invalidate his current claim that $\mathit{NP}=P$. Then in 
Section \ref{sec:Counter-example}, we provide a sound counter-example 
which demonstrates his algorithm does not correctly 
enumerate all perfect matchings in any bipartite graph. In order to 
avoid confusion, our definitions, theorems, and lemmas are labeled 
contiguously, without regard to the section in which they appear. 
Unless noted otherwise, it can be assumed that all other theorems 
and lemmas are from Aslam's paper \cite{aslam}.

We would also like to note that we are critiquing version 9 of Aslam's
paper. Although at time of writing this critique, Aslam has released
two additional versions, 10 and 11, both of those versions are simply
summaries of version 9. As a result, they rely heavily on the claims
made in version 9 and thus it is sufficient to analyze version
9. Whenever we cite Aslam as \cite{aslam}, we refer to version 9 of
his paper.

Aslam represents perfect matchings in bipartite graphs as
permutations. These permutations are elements of the symmetric group
$S_{n}$, the group of all permutations of $n$ elements. A review of
general group theory can be found in any introductory abstract algebra
book such as \cite{fraleigh}.

\subsubsection*{Perfect Matchings as Permutations}

A perfect matching in a bipartite graph $BG_{n}$ with vertices $V\cup W$
is a set of edges represented as

\[
\bigcup_{i=1}^{n}ij,\]

where each $ij$ represents the edge $v_{i}w_{j}$ in $BG_{n}$ with
each $w_{j}\in W$ and $v_{i} \in V$ occuring exactly once and $|V|=|W|=n$. We can
also represent a perfect matching as a permutation
$\pi=(a_{1}, a_{2}, \dots, a_{n})$,
$1\le a_{r} \le n$ for all r; in other words, every element in the
permutation cycle is can be represented as number between $1$ and $n$,
inclusive. Letting $a_{r}^{\pi}=a_{r+1}$ and $a_{n}^{\pi}=a_{1}$,
the perfect matching corresponding to $\pi$ is

\[
E(\pi)=\bigcup_{i=1}^{n}ii^{\pi}.\]

For instance, the permutation $(2,3,1)$ corresponds to the perfect
matching $\{12,23,31\}$, and $(2,3,1,5,4)$ corresponds to $\{15,23,31,42,54\}$.
Futhermore we can decompose each permutation $\pi$ into a product
of transpositions $\pi=\psi_{n}\psi_{n-1}\dots\psi_{1}$ where $\psi_{i}$
is the transposition $(i,k)$ and $i\le k\le n$. From the previous
example, $(2,3,1,5,4)=(5,5)(4,5)(3,5)(2,3)(1,5)$. Note that every
unique $\psi_{n}\psi_{n-1}\dots\psi_{1}$ represents a unique permutation,
i.e., a unique perfect matching for all $n!$ perfect matchings in
a complete bipartite graph.

\subsubsection*{Perfect Matchings from the Generating Graph
  \mbox{\boldmath $\Gamma(n)$}}

A generating graph $\Gamma(n)$ is a DAG (directed acyclic graph)
defined by Aslam to have $O(n)$ vertices called \emph{nodes}. Each
node contains a unique pair of edges $(ik,ji)$ such that $1\le i<k,j\le n$
or $1\le i=j=k\le n$. The graph $\Gamma(n)$ is designed so that
traversal of special paths called complete valid multiplication paths,
$\mathit{CVMP}$s \cite[Definition 4.33]{aslam}, will enumerate every $n!$ perfect matching in a bipartite
graph with $2n$ vertices. There are two ways to find the corresponding
perfect matching given a $\mathit{CVMP}$. If $p=a_{1}a_{2}...a_{n}$ is a
$\mathit{CVMP}$ in $\Gamma(n)$, then one way to find the perfect matching
is from the permutation $\pi$ given by

\[
\pi=\psi(a_{n})\psi(a_{n-1})\dots\psi(a_{1}).\]

where $\psi(a_{i})=\psi(ik,ji)=(i,k)$. The second way to find a perfect
matching is through the union of all edge pairs in each $a_{i}$ without
the surplus edges ($\mathit{SE}$), which will be $jk$ from each edge $a_{i}$$a_{j}$
in $p$, which written formally is
\begin{eqnarray*}
E(p) & = & \ep(p)-(SE(p)\cap EP(p))\\
 & = & \left(\uep\right)-\left(\left(\use\right)\bigcap\left(\uep\right)\right)\end{eqnarray*}
where $\ep(ik,ji)=\{ik,ji\}$ and $\mathit{SE}((ik,ji)(jk,qj))=\{jk\}$, $i<k,j<q$
or $i<k=j=q$. As an example, consider the following $\mathit{CVMP}$ from
the graph $\Gamma(9)$ in Figure \ref{figure1}:
\begin{center}
\begin{figure}[H]
\includegraphics[scale=0.65]{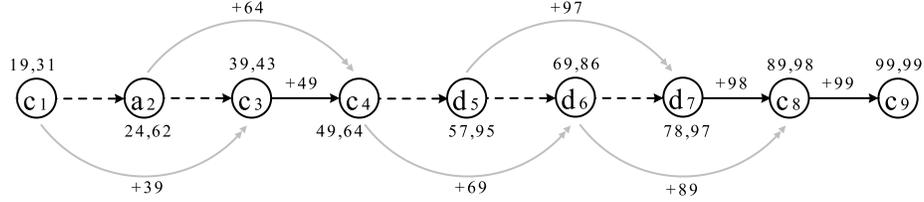}
\caption{\label{figure1}A subgraph of $\Gamma(9)$ containing a $\mathit{CVMP}$ along the dark single-arrowed
edges. Dotted edges are refered to as \emph{S-edges}, solid edges
are \emph{R-edges}, and the edges between nonadjacent nodes with double-arrows
are \emph{jump edges}. Each node is labeled with a pair of edges $(ik,ji)$.
Surplus edges along the $\mathit{CVMP}$ are indicated with a $+jk$ above
them.}
\end{figure}
\end{center}
A $\mathit{CVMP}$ can only be along $S$-edges or $R$-edges which are not
jump edges, so the $\mathit{CVMP}$ pictured above is $p=c_{1}a_{2}c_{3}c_{4}d_{5}d_{6}d_{7}c_{8}c_{9}$.
For each node we can easily find $\psi_{i}$, e.g. $\psi(c_{1})=\psi(19,31)=(19)$,
giving the permutation
\begin{align*}
\pi 
 & =(99)(89)(78)(69)(57)(49)(39)(24)(19)\\
 & =(1,9,5,7,8,6,2,4,3).
\end{align*}
We can also easily compute $E(p)$ from the set of all edges in all
the nodes minus all of the surplus edges as
\begin{align*}
E(p) & = \ep(c_{1})\cup\ep(a_{2})\cup...\cup\ep(c_{9})-SE(c_{1})\cup SE(a_{2})\cup...\cup SE(c_{9})\\
 & = \{19,31,24,62,...,98,99\}-\{39,64,...,99\}\\
 & = \{19,24,31,43,57,62,78,86,95\}.
\end{align*}
Clearly, $E(\pi)=E(p)$ so both methods have arrived at the same perfect
matching.

\subsubsection*{\mbox{\boldmath $\mathit{CVMP}$}s in Incomplete Graphs}

Each of the $n!$ $\mathit{CVMP}$s in $\Gamma(n)$ corresponds to a unique
perfect matching in a complete bipartite graph $BG_{n}$. Of course,
given an incomplete bipartite graph $\bg$ only a subset of all the
$\mathit{CVMP}$s will contain edges not in $\bg$ which cannot be counted as a perfect
matchings. These edges not in $\bg$ are refered to as the edge requirements
($\mathit{ER}$) of a $\mathit{CVMP}$ and are formally defined for a $\mathit{CVMP}$ $p$ as
\begin{eqnarray*}
ER(p) & = & \left(\uer \right)-\\
& & {} \left(\left(\use\right)\bigcap\left(\uer\right)\right),
\end{eqnarray*}
where $\mathit{ER(a_{i})}=\{e|e\in\ep(a_{i}),e\notin\bg\}$. Clearly, the $\mathit{ER}$
of a $\mathit{CVMP}$ will be the empty set if and only if the $\mathit{CVMP}$ corresponds to
a valid perfect matching in $\bg$. In other words, $\mathit{ER(p)}=\emptyset\iff E(p)\subseteq\bg$.
Performing a \emph{valid enumeration} of $\mathit{CVMP}$s in $\Gamma(n)$
given a corresponding $\bg$, i.e. enumerating every $\mathit{CVMP}$ with
$\mathit{ER(p)}=\emptyset$, is equivalent to counting every valid perfect
matching.

Obviously, naive valid enumeration of the $n!$ possible $\mathit{CVMP}$s
cannot be done in polynomial time. However, if Aslam's algorithm is
correct, the special properties of $\Gamma(n)$ allow it to be reduced
to sets of subpaths which can be joined while preserving the $\mathit{ER}$
of all $\mathit{CVMP}$s.

\subsubsection*{Algorithm Overview}

A generating graph $\Gamma(n)$ contains $O(n^{3})$ nodes \cite[Property 4.21]{aslam} and can
be created in polynomial time. $\vs(a_{i},a_{j})$ is a data structure
representing a subset of $\mathit{VMP}$s between nodes $a_{i}$ and $a_{j}$
that have the same $\mathit{ER}$. Within the algorithm, all of the $\vs(a_{i},a_{i+1})$
are initializated first and all possible $\vs(a_{i},a_{j})$ are stored
in a matrix.

During each iteration, the algorithm performs two reduction operations:
adding and multiplying $\vs$s. If one multiplies two $\vs$s, $\vs(a,b)$ and
$\vs(b,c)$, effectively doubles the length of the paths they represent
to get $\vs(a,c)=\vs(a,b)\times\vs(b,c)$, and increases the number
of $\mathit{VMP}$s in the new $\mathit{VMPSet(a,c)}$ to \[
|\vs(a,c)|=|\vs(a,b)|\times|\vs(b,c)|.\]
When two sets $\vs(a,b)$ and $\vs^{\prime}(a,b)$ can be combined
so as to satisfy conditions for multiplication they are added together,
and the number of $\mathit{VMP}$ in the new $\vs^{\prime\prime}(a,b)=\vs(a,b)\cup\vs^{\prime}(a,b)$
is

\[
|\vs^{\prime\prime}(a,b)|=|\vs(a,b)|+|\vs^{\prime}(a,b)|.\]

Since every iteration doubles the length of $\mathit{VMP}$s represented in
the $\vs$s, after $O(\log(n))$ iterations the entire generating
graph $\Gamma(n)$ is reduced to several disjoint sets of $\vs(1,n)$
representing all $\mathit{CVMP}$s. Summation over all $|\vs(1,n)|$ that have
$\mathit{ER}=\emptyset$ then gives the total number $\mathit{CVMP}$s and likewise
perfect matchings. The summation step is what necessitates keeping
$\mathit{ER}$ the same for all $\vs$s.

Since the problem of counting perfect matchings in any bipartite graph
is $\#P$-complete and the class of parallel algorithms running in
$(\log n)^{O(1)}$ on a polynomial number of processors, which is
analogous to and commonly referred to as $\mathit{NC}$, is a subset
of $\mathit{FP}$, the algorithm Aslam purports to have will be able to solve
any $\#P$ problem in polynomial time, meaning $\#P=FP$ and $NP=P$.


\section{Refutation} \label{refutations}
The main flaw in Aslam's reasoning is that the $\mathit{ER}$ of every $\mathit{CVMP}$
can be preserved during decomposition and subsequent reduction operations
over $\vs$s. However, the $\mathit{ER}$ of a $\vs$ does not capture the
$\mathit{SE}$ of the $\mathit{VMP}$s it contains, and ultimately $\mathit{SE}$ will determine
$\mathit{ER}$. We will show that multiplication and addition of $\vs$s does
not always give a $\vs$ with the same $\mathit{ER}$ for all $\mathit{VMP}$s. As proof
this problem is inherent in all sufficiently large generating graphs
we give a counter-example in Section \ref{sec:Counter-example}.

\begin{lem_s} \label{lemma_ref_g1}
The product $\vs$ $\mathit{AC}$ found from multiplying two $\vs$s
$A=\vs(a,b)$ and $C=\vs(b,c)$ must be a single set and contain only
$\mathit{VMP}$s with the same $\mathit{ER}$.
\end{lem_s}
\begin{proof}
The condition on $\mathit{ER}$ follows from the definition
of $\vs$ and the inductive result of the algorithm itself. Since
after the final iteration all $\mathit{VMPSet}$s with $\mathit{ER}=\emptyset$ will
be counted, if multiplication resulted in a $\mathit{VMPSet}$ containing some
$\mathit{CVMP}$s with $\mathit{ER}=\emptyset$ (valid perfect matchings) and some with
$\mathit{ER}\ne\emptyset$ (invalid), then summation would result in an incorrect
number of perfect matchings.
\end{proof}
By Aslam's definition of multiplication, $\mathit{AC}$ must be a
single $\vs$. We note, however, even if we allowed multiplication to produce more
than one $\mathit{VMPSet}$, the number of sets produced would have to
be constant with respect to $|A|\times|C|$, since multiplication
is what allows the fast enumeration in $O(\log(n))$ iterations.
\begin{thm} \label{theorem_ref_g1}
The conditions for multiplying $\vs$s $A\times C=AC$ given
in Lemma 5.8 and Lemma 5.9 of Aslam's proof are insufficient
conditions for multiplication. Lemma 5.9 is not a necessary
condition for multiplication.
\end{thm}

\begin{proof}
Lemma 5.8 states that $\forall p\in A$ and $\forall q\in C$, $p$
must multiply (form a $\mathit{VMP}$) with $q$. Obviously, this is a necessary
condition since we are considering all $pq$ to be $\mathit{VMP}$s in $AC$.
Lemma 5.9 states that every node covered by a $\mathit{VMP}$ in $C$ from
the same partition must have the same $\mathit{ER}$. Note that partition number
is equal to the depth in $\Gamma(n)$, so for every $x_{i},x_{i}^{\prime}\in q$
we must have $\mathit{ER(x_{i})}=\mathit{ER(x_{i}^{\prime})}$. The proof of this lemma
is conspicuously omitted and we found it to be incorrect. Assume the
only nodes with unequal $\mathit{ER}$ in $C$ are $x_{i}\in q$ and $x_{i}^{\prime}\in q^{\prime}$,
all the $\mathit{VMP}$s in $A$ have the same $\mathit{SE}$, and let $\mathit{ER(x_{i})}=e$
and $\mathit{ER(x_{i}^{\prime})}=\emptyset$. If $e\notin \mathit{SE}(A)$ the resultant
$AC$ will not have every $\mathit{VMP}$ with the same $\mathit{ER}$ violating the
condition laid out in Lemma 1. However, if $e\in SE(A)$ then $\mathit{ER(pq^{\prime})}=\mathit{ER}(pq)=\emptyset$
and $\mathit{AC}$ is a $\mathit{VMPSet}$ with valid $\mathit{VMP}$s all with $\mathit{ER}=\emptyset$,
so Lemma 5.9 is not a necessary condition for multiplication. Note
we use the following definitions:
\begin{eqnarray*}
e\in SE(A) & \iff & \left(\forall p\in A\right)\,\left[e\in SE(p)\right].\\
e\in ER(A) & \iff & \left(\forall p\in A\right)\,\left[e\in ER(p)\right].\end{eqnarray*}

Now we will show these lemmas are insufficient conditions for multiplication.
Let $p$ and $p^{\prime}$ be $\mathit{VMP}$s in $A$ such that $e\in \mathit{SE(p)}$
and $e\notin \mathit{SE(p^{\prime})}$. Let the node $x_{i}$ covered by all
$\mathit{VMP}$s in $C$ have $\mathit{ER}=e$ and all other nodes have $\mathit{ER}=\emptyset$.
Note that $C$ satisfies Lemma 5.9 since every $\mathit{VMP}$ in $C$ covers
$x_{i}$ there is no node $x_{i}^{\prime}$ in $C$ and every other
node has the same $\mathit{ER}=\emptyset$. Proof that $A$ and $C$ can satisfy
Lemma 5.8 (every path through $a$ and $c$ is a $\mathit{VMP}$) relies on
properties of the generating graph $\Gamma(n)$ itself and so we will
demonstrate that in our counter-example in Section \ref{sec:Counter-example}. For now we assume
Lemma 5.8 is satisfied.

After multiplication of $A$ and $C$, we have $\mathit{ER(pq)}=\emptyset$
and $\mathit{ER(p^{\prime}q)}=e$ for every $q\in C$. The resultant $\vs$
$AC$ does not contain $\mathit{VMP}$s with the same $ER$, violating the
conditions for multiplication set forth in Lemma 1. We see this is
because the $\mathit{VMPSet}$ representation does not capture cases where
$\mathit{SE}(p)\ne \mathit{SE}(p^{\prime})$.
\end{proof}

\subsubsection*{Further Discussion}

Although Aslam does not give the sufficient conditions for performing
multiplication to reduce all $\vs$s in $O(\log(n))$ iterations,
if his proof is valid until that point, then proving whether
such conditions exist or do not exist may be equivalent to proving
whether $P=NP$. In the rest of the section we explore why performing
multiplication while preserving $\mathit{ER}$ over $\vs$s is difficult. In
the following lemma we show why the various $\mathit{SE}$s of all the paths
in a $\vs$ are significant.

\begin{lem_s}
There are at least $(n-1)!$ $\mathit{CVMP}$s where $\mathit{SE}(p)\subsetneq EP(p)$
for every $\mathit{CVMP}$ $p$. In these $\mathit{CVMP}$s, if $a_{i}\in p$ and all
edges $\mathit{SE(a_{i})}$ are not present in the bipartite graph $\bg$,
then changing any one node in $p$ results in $\mathit{ER(p)}\ne\emptyset$.
\end{lem_s}

\begin{proof}
Each $\mathit{CVMP}$ $p$ in $\Gamma(n)$ is $n$ nodes long.
For simplicity, we only consider the $(n-1)!$ cases where $p$ is
composed of non-identity nodes (except the last node). Recall
\begin{eqnarray*}
E(p) & = & \ep(p)-\left(SE(p)\cap EP(p)\right)\\
 & = & \left(\uep\right)-\left(\left(\use\right)\bigcap\left(\uep\right)\right).\end{eqnarray*}
Every node $a_{i}=(ik,ji)$, $1\le i<j,k<n$ contributes two unique
edges ${ik,ji}$ to $EP(p)$ and the last node contributes one so
that $|\ep(p)|=2n-1$. Every node except the last contributes an edge
to $\mathit{SE}(p)$ so $|SE(p)|=n-1$. Since $|E(p)|=n$, every edge in $\mathit{SE(p)}$
must be unique in $p$, eliminate one edge from $\ep(p)$, and $\mathit{SE(p)}\subsetneq \mathit{EP(p)}$.
Note that each $a_{i}=(ik,ji)\in\Gamma(n)$ is unique, so for all
nodes $\mathit{SE}(a_{i})\ne SE(b_{i})$. Therefore if the edge $\mathit{SE}(a_{i})$
is not present in the bipartite graph $\bg$, then changing the node
$a_{i}\in p$ to any $b_{i}\in\Gamma(n)$ results in an $\mathit{ER(p)}\ne\emptyset$,
where

\begin{eqnarray*}
ER(p) & = & \left(\uer\right)-\\
& & {} \left(\left(\use\right)\bigcap\left(\uer\right)\right).\end{eqnarray*}

It follows there are at least $(n-1)!$ $\mathit{CVMP}$s with corresponding
$\bg$s in which satisfying $\mathit{ER}=\emptyset$ may be dependent on the
$\mathit{SE}$ of every node in $p$. If we are concerned with satisfying the
conditions on multiplication outlined in Lemma \ref{lemma_ref_g1}, we can keep all
$\mathit{VMP}$s with differing $\mathit{SE}$ in separate $\vs$s, but this leads to
a large number of $\vs$s.
\end{proof}

\begin{lem_s}
The number of $\vs$s with the same $\mathit{SE}$ from partition $1$
to $i$ is at least $\binom{n}{i}$.
\end{lem_s}
\begin{proof}
By definition, the generating graph $\Gamma(n)$ contains $n!/(n-i)!$
$\mathit{VMP}$s from partitions $1$ to $i$. Note this is necessary for $\Gamma(n)$
to be able to enumerate $n!$ perfect matchings. Since all these $\mathit{VMP}$s
are unique, for any two $\mathit{VMP}$s $p_{1}$ and $p_{2}$, each will contain
at least one different node $x_{i}\in p_{1}$ and $y_{i}\in p_{2}$,
with $\mathit{SE(a_{i})}\ne \mathit{SE(b_{i})}$. Consequently, no two $\mathit{VMP}$s have
surplus edges occuring in exactly the same order, so the maximum size
of any $\vs$ with the same set of $\mathit{SE}$ edges from $1$ to $i$ will
be $i!$ which is the number of permutations of $\mathit{SE}$ edges from $i$
nodes. To get a lower bound on the number of $\vs$s with the same
$\mathit{SE}$ we divide the number of $\mathit{VMP}$s by the upper bound on set size:

\[
\dfrac{n!}{(n-i)!i!}=\binom{n}{i}. \qedhere \]
\end{proof}
\section{Counter-example\label{sec:Counter-example}}

We present $\gamma$, a subgraph of every $\Gamma(n)$ with $n\ge9$,
which enumerates five perfect matchings. We then offer an incomplete
bipartite graph $\bg$ and show that Aslam's algorithm will incorrectly
count some number of perfect matchings in $\bg$ using $\gamma$. This
serves as an example that the graph described in Theorem \ref{theorem_ref_g1} can be realized in $\Gamma(n)$, satisfying Lemma 5.8 of Aslam's paper.

\begin{center}
\begin{figure}[t]
\includegraphics[scale=0.65]{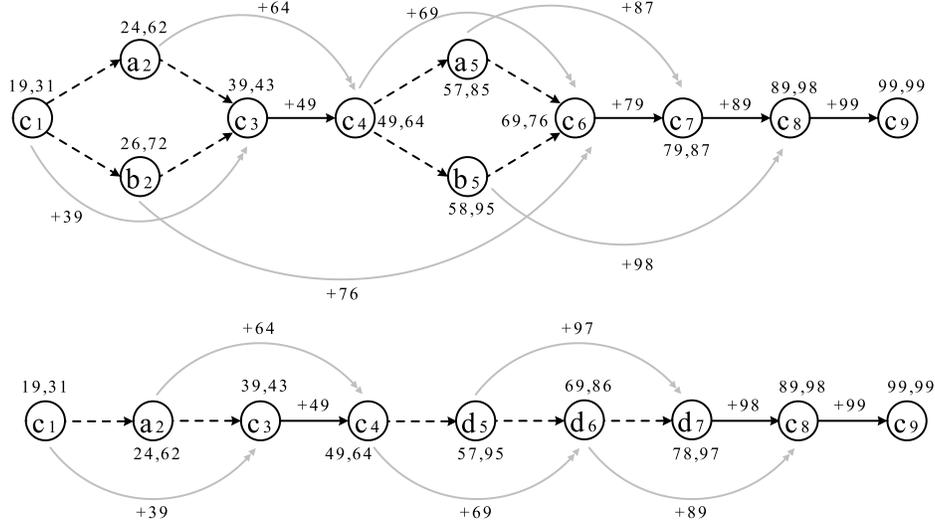}
\caption{\label{figure2} The above figure represents graph $\gamma$. Double-arrowed solid-lines are jump
edges, single-arrowed solid-lines are R-edges, and dotted-lines are
S-edges.}
\end{figure}
\end{center}

\paragraph{Graphical Representation of the Example}
In Figure \ref{figure2} we provide a graphical representation of the counter-example.

\begin{lem_s}
$\gamma$, in Figure \ref{figure2}, represents a subgraph of every $\Gamma(n)$, where
$n\ge9$.
\end{lem_s}

\begin{proof}
The generating graph $\Gamma(n)$ is defined over all $1\le i<k\le n$,
where vertices are $a_{i}\in g(i)$, $R$-edges are $a_{i}a_{j}$
such that $|a_{i}Ra_{j}|=1$, jump edges are $R$-edges such that
$j\ne i+1$, and $S$-edges are $a_{i}a_{i+1}$ such that $a_{i}Sa_{i+1}$
\cite[Definition 4.12]{aslam}. From this definition we should note that the generating
graph $\Gamma(n)$ is a subgraph of $\Gamma(n+1)$.

Now we will show that $\gamma$ from is a valid subgraph of $\Gamma(9)$.
It can be easily verified every node $a_{i}=(ik_{1},k_{2}i)$ in $\gamma$
has either $k_{1},k_{2}>i$ or $i=k_{1}=k_{2}$, so $a_{i}\in g(i)$
is true for all nodes. Any relation $R$ in $\gamma$ between nodes
$a_{i}$ and $a_{j}=(jt_{1},t_{2}j)$ appears iff $t_{1}=k_{2}$,
$t_{2}=k_{1}$, and $i<j$, which satisfies $|a_{i}Ra_{j}|=1$. The
relation appears as an $R$-edge if $j=i+1$, otherwise it appears
as a jump-edge. Any $S$-edge appears if and only if $j=i+1$, and either $k_{1},t_{2}<k_{2},t_{1}$,
$k_{2},t_{1}<k_{1},t_{2}$, or $k_{1}=k_{2}<t_{1},t_{2}$. The disjointness
of $a_{i}Sa_{j}$ given these conditions is made clear from Remark
4.5 \cite{aslam} since no $R$-cycle $C_{a_{i}a_{j}}$can be constructed which
has a strictly increasing or decreasing traversal.

Since for all nodes in $\gamma$ we have $a_{i}\in g(i)$, and all
edges appear if and only if they satisfy their respective definitions (at least
for the nodes appearing in $\gamma$), $\gamma$ is a subgraph of
$\Gamma(9)$ and every $\Gamma(n)$, where $n\ge9$.
\end{proof}

\begin{cvmps_in_gamma}
Let $P(a,b)$ denote the path in $\gamma$ starting at $c_{1}$
going through $a,b$ and ending at $c_{9}$. Then $\gamma$ only contains
five $\mathit{CVMP}$s: $p_{aa}=P(a_{2},a_{5})$, $p_{ab}=P(a_{2},b{}_{5})$,
$p_{ba}=P(b_{2},a_{5})$, $p_{bb}=P(b_{2},b_{5})$, and $p_{ad}=P(a_{2},d_{5})$
which correspond to five unique perfect matchings.
\end{cvmps_in_gamma}

\begin{proof}
By definition the permutation $\pi(p)$ that realizes the perfect
matching corresponding to $\mathit{CVMP}$ $p$ is given by $\psi_{n}\dots\psi_{1}$,
where for every node $x_{i}=(ik,ji)\in p$, $\psi_{i}=(ik)$. The
set of edges in the perfect matching corresponding to $\mathit{CVMP}$ $p$
is given by $E(p)=\left(\bigcup\{e|e\in x_{i}\}\right)-\left(\bigcup SE(x_{i})\right)$, where
$\mathit{SE}(x_{i})$ gives the surplus edges in $p$.

It is trival to verify $\pi(p)$ and $E(p)$ are consistent:
\begin{center}
$\pi(p_{aa})=(99)(89)(79)(69)(57)(49)(39)(24)(19)$
$E(p_{aa})=\{19,24,31,43,57,62,76,85,98\}$
\end{center}
\begin{center}
$\pi(p_{ab})=(99)(89)(79)(69)(58)(49)(39)(24)(19)$
$E(p_{ab})=\{19,24,31,43,58,62,76,87,95\}$
\end{center}
\begin{center}
$\pi(p_{ba})=(99)(89)(79)(69)(57)(49)(39)(26)(19)$
$E(p_{ba})=\{19,26,31,43,57,64,72,85,98\}$
\end{center}
\begin{center}
$\pi(p_{bb})=(99)(89)(79)(69)(58)(49)(39)(26)(19)$
$E(p_{bb})=\{19,26,31,43,58,64,72,87,95\}$
\end{center}
\begin{center}
$\pi(p_{ad})=(99)(89)(78)(69)(57)(49)(39)(24)(19)$
$E(p_{ad})=\{19,24,31,43,57,62,78,86,95\}.$
\end{center}
Note that $P(b_{2},d_{5})$ is not a $\mathit{CVMP}$ because it does not contain
the mdag $MDG(b_{2},c_{3},c_{6})$.
\paragraph{Choosing \mbox{\boldmath $BG^{\prime}$}}

The initialization of $\mathit{PTM}$, which Aslam defines as the matrix containing all $\vs$s, is
vague regarding how it incorporates information from the adjacency
matrix $\mathit{BGX}$ of the bipartite graph, considering there is no clear
relation between removal of $S$-edges or $R$-edges, and edges from
the bipartite graph. To allow this ambiguity, we have chosen a $BG^{\prime}$
so that no edge might be removed from $\gamma$ without losing a perfect
matching. Formally,

\[
BG^{\prime}=\left(\bigcup E(p)\right)-\{76\}\]

In other words, all $\mathit{CVMP}$s in $\gamma$ are valid matchings with
$\mathit{ER(p)}=\emptyset$ except for those containing the edge $\{76$\}
which will have $\mathit{ER(p)}=\{76\}$. Clearly only $p_{ba}$, $p_{bb}$,
and $p_{ad}$ are $\mathit{CVMP}$s which represent valid perfect matchings
so the result of enumerating with $\gamma$ should be 3. In addition,
these $\mathit{CVMP}$s cover every edge of $\gamma$, so no edge may be removed
without losing at least one $\mathit{CVMP}$.

\subsubsection*{Iterations of Add and Join on \mbox{\boldmath $\gamma$}}

During the first iteration of the algorithm, the $\vs$s $\vs A(c_{1},c_{3})$,
$\vs B(c_{1},c_{3})$, $\vs A(c_{4},c_{6})$, $\vs B(c_{4},c_{6})$,
and $\vs$s over all other pairs of nodes are created. In the next
iteration two additions between the corresponding $A$ and $B$ sets occur:
$\vs A(c_{1},c_{3})$ and $\vs B(c_{1},c_{3})$ are added
together, and $\vs A(c_{4},c_{6})$ and $\vs B(c_{4},c_{6})$ are
added. So we have
\begin{eqnarray*}
\vs(c_{4},c_{6}) & = & \vs A(c_{4},c_{6})+\vs B(c_{4},c_{6})\\
\vs(c_{1},c_{3}) & = & \vs A(c_{1},c_{3})+\vs B(c_{1},c_{3}).\end{eqnarray*}
Crucially, $|\vs(c_{4},c_{6})|=|\vs(c_{1},c_{3})|=2$ and the set of surplus edges
$\mathit{SE}$ of $\vs(c_{4},c_{6})$ equals the $\mathit{ER}$ of $\vs(c_{1},c_{3})$ which
is $\{76\}$. Once these two sets are multiplied to get 4 and the
additional $\vs$ representing $p_{ad}$ is counted, the number of
perfect matchings the algorithm will return will be 2 more than the
actual number of perfect matchings because the $\mathit{SE}$ of the $\mathit{VMP}$s
in $\vs(c_{4},c_{6})$ was combined.
\end{proof}
Thus, as the above example demonstrates, Aslam's algorithm does not
correctly enumerate all perfect matchings for all cases. Therefore,
his current proof that $\#P \subseteq FP$ (and hence that $P=NP$ and the
polynomial-time hierarchy collapses) does not validly establish that claim.
\section{Acknowledgements}
This work was completed in partial fulfillment of the
requirements for an Honors Bachelor of Science Degree in Computer
Science from the Department of Computer Science at the University of
Rochester, in Rochester, NY, USA. This paper was also written as the
Honors Project for the course CSC200H during the Spring 2009
semester. We would like to thank Professor Lane A. Hemaspaandra, the
course T.A. Adam Sadilek and others in the community for their
feedback, support and suggestions.

\end{document}